\documentstyle[12pt]{article}

\newcommand{\ba}{\begin{array}}
\newcommand{\ea}{\end{array}}
\newcommand{\be}{\begin{equation}}
\newcommand{\ee}{\end{equation}}
\newcommand{\bea}{\begin{eqnarray}}
\newcommand{\eea}{\end{eqnarray}}
\newcommand{\beas}{\begin{eqnarray*}}
\newcommand{\eeas}{\end{eqnarray*}}

\def\wt{\widetilde}

\def\sect{\section}
\def\EQ{\begin{equation}}
\def\EN{\end{equation}}
\def\bea{\begin{eqnarray}}
\def\ena{\end{eqnarray}}
\newcommand{\vs}[1]{\vspace{#1 mm}}

\newcommand{\mathbb}[1]{{\bf{#1}}}

\begin{document}
\begin{titlepage}
\begin{center}

\hfill  {\sc September 1999} \\
\hfill  {\tt hep-th/9909081} \\
        [.35in]

{\large\bf 
$D$--brane Physics and Noncommutative Yang--Mills Theory }\\
\medskip
\medskip

{\bf Lorenzo Cornalba}\footnote{E-mail: {\tt cornalba@princeton.edu} \\ \indent
Address after October 1st, 1999: {\it I.H.E.S., Le Bois-Marie, Bures-Sur-Yvette, 
91440 France}} \\
{\it Department of Physics}\\
{\it Joseph Henry Laboratories, Princeton University}\\
{\it Princeton, NJ 08544, U.S.A.}\\

\end{center}

\vs{4}
\centerline{{\bf{Abstract}}}
\vs{4}

We discuss the physics of a single $Dp$--brane in the presence of a background 
electromagnetic field $B_{ij}$.
It has recently been shown \cite{SW} that, in a specific $\alpha '
\rightarrow 0$ limit, the physics of the brane is correctly described by
noncommutative Yang--Mills theory, where the noncommutative gauge potential is given 
explicitly in terms of the ordinary $U(1)$ field. In a previous paper \cite{SC}
the physics of a $D2$--brane was analyzed in the Sen--Seiberg limit of M(atrix) theory 
by considering a specific coordinate change on the brane world--volume.

We show in this
note that the limit considered in \cite{SC} is the same as the one described in \cite{SW},
in the specific case $p=2$, ${\rm rk}\ B_{ij} = 2$. Moreover we show that the coordinate change in
\cite{SC} can be reinterpreted, in the spirit of \cite{SW}, as a field redefinition of
the ordinary Yang--Mills field, and we prove that the transformations agree for large 
backgrounds.
The results are finally used to considerably streamline the proof of the equivalence of the 
standard Born--Infeld action with noncommutative Yang--Mills theory, in 
the large wave--length regime.

\end{titlepage}

\newpage
\renewcommand{\thefootnote}{\arabic{footnote}}
\setcounter{footnote}{0}

\sect{Introduction}

In this short note we are going to examine the physics of a single flat $D$--brane in 
the presence of a large background electromagnetic field. The problem has received lately 
considerable attention and has been analyzed in complete detail in the recent 
work \cite{SW}. In particular it has been shown in \cite{SW} that, in order to describe the
physics
of the fluctuating electromagnetic
field, one has two equivalent options. On one side, one can follow
the standard treatment of the subject in terms of an ordinary
$U(1)$ gauge field in the presence of the background field strength. On the other side,
one can fully include the effects of the background by rewriting the action in terms
of a $U(1)$ gauge field of noncommutative gauge theory. In this formulation,
the background is used to define a Poisson structure and a 
star product on the brane world--volume, which are then used to define the noncommutative
gauge theory. Ordinary gauge transformations, which form an abelian group, are 
replaced by noncommutative gauge transformations, which give rise to a non abelian
group. Nonetheless gauge equivalent configurations are so in both descriptions, and
therefore the gauge orbits are the same in both cases. Finally, in \cite{SW}, the authors 
identify a specific $\alpha ' \rightarrow 0$
limit in which the noncommutative description simplifies 
considerably and reduces to standard noncommutative Yang--Mills theory.

In a previous work  \cite{SC}, a similar setting and limit
was considered in the analysis of the physics of $D2$--branes in Type IIA string theory.
The motivations for the analysis in \cite{SC} are quite different from the ones in \cite{SW},
and the discussion is driven by the attempt to connect the physics of a $D2$--brane to
that of an $M2$--brane in the $11$--dimensional strong coupling limit of Type IIA.
In particular, in order to connect with the $11$--dimensional 
interpretation of the theory, it was crucial in \cite{SC}
to treat on the same footing the directions
transverse and parallel to the brane. This can be achieved by changing coordinates on the
world--volume of the brane, so as to eliminate any fluctuation of the field strength
in favor of fluctuations of the induced metric. Since the action governing the brane is
invariant under diffeomorphisms, the coordinate change can be considered as a field
redefinition, and does not alter the physics. On the other hand, in terms of the new
fields, the action for a $D2$--brane possesses a smooth polynomial limit in the 
$11$--dimensional limit of Type IIA. 

In this note we will show that the limit considered in \cite{SC} is that of \cite{SW} in the
specific case of a $D2$--brane. Moreover, we will show that the coordinate change 
considered in \cite{SC} is the same, to leading order in derivatives, to the one considered in 
\cite{SW}. The result can be understood as follows. As we have briefly described above, and as
we shall explain in detail later, it is convenient to change coordinates in order
to eliminate
any fluctuations of the $U(1)$ field strength. This change of coordinates in not
unique, but is defined up to diffeomorphisms of the world--volume of the brane which leave
invariant the background field strength $2$--form. The group of such diffeomorphisms
is non abelian, and has as infinitesimal generators the Hamiltonian vector fields
defined in terms of the background, now considered as a symplectic structure on the
world--volume. These diffeomorphisms replace the ordinary abelian gauge invariance of the
original theory, as in the case of \cite{SW}. Moreover, to leading order in derivatives, 
commutators in terms of the star product considered in \cite{SW} are nothing but Poisson 
brackets with respect to the symplectic structure defined by the background. This explains
why the change of coordinates is the same to leading order. 

Given the above facts, we show that the results of \cite{SC}, generalized to the case of a 
generic $Dp$--brane, can be used to streamline the proof of the equivalence of the 
standard Born--Infeld action with noncommutative Yang--Mills theory, in the $\alpha '
\rightarrow 0$ limit considered in \cite{SW}. Moreover, in the large wave--length regime,
the results in this note give a clear geometric interpretation to the field redefinition
given in \cite{SW}.

We are going to use primarily the notation of \cite{SW}, in order to allow a quick comparison 
of the equations. Moreover, given the nature of this short note,
we are not including an extensive reference list. For a more complete bibliography we refer the 
reader to the papers \cite{SW} and \cite{SC}.

\sect{The Coordinate Redefinition}

In this section we are going to analyze the physics of a flat $Dp$--brane in Type II string
theory, whose 
world--volume is extended in the spacetime directions $0,\cdots,p$.
We are not going to consider
the transverse motions of the brane, and therefore we will limit our attention 
exclusively to the degrees of freedom of the $U(1)$ gauge potential living on the brane 
world--volume. In particular we will analyze the dynamics of the $Dp$--brane in the 
presence of a large background electromagnetic field.

We will denote with $X^i$, $i=0,\cdots,p$, the coordinates on the target 
spacetime which are parallel to the $Dp$--brane. The world--volume $\Sigma = {\bf R}^{p+1}$ 
of the brane can be parameterized
with the natural coordinates $x^i$ $(i=0,\cdots,p)$ inherited from the target,
which are defined
by 
$$
X^i (x) = x^i.
$$
The embedding functions are therefore non--dynamical, and all of the physics of the $Dp$--
brane (recall that we are neglecting transverse motions) is described by the fluctuating
$U(1)$ gauge potential. As we already mentioned, we are going to work in the presence of a 
large constant background magnetic field $B_{ij}$. We will assume in this paper 
that the constant matrix $B$ is of maximal rank. For the most part of what follows, we shall
actually assume that
$p$ is odd, and therefore that $B$ is invertible (${\rm rk}\ B = p+1$).
Following the notation of \cite{SW}, we shall in this case
denote with $\theta = B^{-1}$ the inverse matrix.
At the end of the section we will briefly return to the case of even $p$ (${\rm rk}\ B = p$).

Let us denote with $a = a_i (x) dx^i$ the 
fluctuating gauge potential, and with $f = da$ the corresponding field strength. The total
field strength is then given by
\be\label{eq300}
F_{ij} (x) =  B_{ij} + f_{ij} (x) = B_{ij} + \partial_i a_j (x) - \partial_j a_i (x).
\ee

We now change parameterization on the world--volume $\Sigma$ by choosing new coordinates 
$\sigma^i$. Clearly the target--space embedding coordinates are now given by
$$
X^i(\sigma) = x^i(\sigma).
$$
Moreover the new field strength is now given by
\be\label{eq100}
\wt F_{ij}(\sigma) = \frac{\partial x^k}{\partial \sigma^i}
\frac{\partial x^l}{\partial \sigma^j} F_{kl} (x(\sigma)).
\ee
We claim that we can choose the coordinates $\sigma$ so that the the field strength 
$\wt F_{ij}$ is given by
\be\label{eq200}
\wt F_{ij} (\sigma) = B_{ij}.
\ee
In other words, we can eliminate any fluctuation of the electromagnetic field by a 
simple coordinate redefinition. In the new coordinate system $\sigma^i$ the dynamics
of the brane is not described by the $U(1)$ gauge potential $a_i(x)$, but is now 
equivalently described by the embedding functions $x^i(\sigma)$, which are now the
dynamical fields.

Let us start our analysis by considering small fluctuations, and therefore by working to 
first order in $a_i$. First let us define the displacement functions
$$
x^i(\sigma) = \sigma^i + d^i(\sigma).
$$
For small fluctuations, $d^i$ is of order $o(a)$.
To first order in fluctuations equation (\ref{eq100}) becomes
$$
\wt F_{ij} = F_{ij} + {\cal L}_d F_{ij} = F_{ij} + d^k \partial_k F_{ij} + F_{kj} 
\partial_i d^k +  F_{ik} \partial_j d^k,
$$
where ${\cal L}_d$ is the Lie derivative in the direction of the vector--field $d^i$.
Using equations (\ref{eq200}) and (\ref{eq300}), and recalling the antisymmetry of $B_{ij}$,
we can rewrite the above (to order $o(a)$) as
$$
f_{ij} = \partial_i a_j - \partial_j a_i 
= B_{jk}\partial_i d^k - B_{ik}\partial_j d^k.
$$
The above is an identity if and only if
\be\label{eq3000}
a_i = B_{ij} d^j + \partial_i \lambda
\ee
for some scalar $\lambda$. Therefore
$$
d^i =  \theta^{ij} a_j
$$
up to transformations of the form
$$
d^i \rightarrow d^i + \theta^{ij} \partial_j\lambda.
$$

Let us now move to the general case. Suppose that we are given a generic gauge field $a_i(x)$,
and a corresponding coordinate change $x^i(\sigma)$ which satisfies (\ref{eq200}). We can 
then analyze, given an infinitesimal change $a_i \rightarrow a_i + \delta a_i$, the
corresponding variation $x^i \rightarrow x^i + \delta x^i$. In fact we can use the results just 
derived above in the linear approximation. First let us change coordinates 
from $x^i$ to $\sigma^i$. The gauge field fluctuation $\delta a_i(x)$ transforms as
$
\delta a_i(x) \rightarrow \wt{\delta a}_i (\sigma) = 
\delta a_j(x(\sigma)) \partial_i x^j
$. We are now in the condition analyzed previously, since, by construction, the 
field strength in the coordinates $\sigma^i$ is equal to $B_{ij}$, and we are considering
an additional infinitesimal change $\wt{\delta a}_i$ to the gauge potential.
We can now use the above results on infinitesimal fluctuations to conclude that
$x^i(\sigma) \rightarrow x^i(\sigma +\eta)$, where $\eta^i(\sigma) =\theta^{ij}
\wt{\delta a}_j(\sigma)$. Therefore $\delta x^i = \eta^j \partial_j x^i$, and 
we have the relation
\be\label{eq2000}
\delta x^i(\sigma) = \theta^{jk}  \partial_j x^i \partial_k x^l\ \delta a_l(x(\sigma)).
\ee
We have seen before that, to first order in $a_i$, $x^i = \sigma^i + \theta^{ij} a_j$.
Using this result and integrating the above equation we conclude that,
to second order in $a_i$, we have
$$
x^i = \sigma^i +\theta^{ij}a_j+ \frac{1}{2}
\theta^{ij} \theta^{kl} \left(2 a_l\partial_k a_j + a_k \partial_j a_l
\right)+\cdots.
$$
     
To make contact with the work of \cite{SW}, let me define the following variable
$$
\hat A_i(\sigma) = B_{ij} d^j(\sigma)
$$
which is given, in terms of $a$, by the formula
\be\label{eq600}
\hat A_i = a_i + \frac{1}{2}
 \theta^{kl} \left(2 a_l\partial_k a_i + a_k \partial_i a_l
\right)+\cdots.
\ee
We clearly see that, to first non--trivial order in $\theta$, the field $\hat A_i$ 
corresponds to the one described in \cite{SW}. We will now better analyze this correspondence
by describing how the original gauge invariance of the theory manifests itself in terms
of the new dynamical fields $d^i$ or, alternatively, $\hat A_i$.

The coordinate change from the variables $x^i$ to the variables $\sigma^i$ is defined so
that the electromagnetic field, in the coordinates $\sigma^i$, is given by the constant 
matrix $B_{ij}$. Clearly the choice of coordinates $\sigma^i$ is not unique. In fact,
given an infinitesimal vector field $V^i(\sigma)$, we may define new 
coordinates $\sigma^i + V^i(\sigma)$ which are equally valid if ${\cal L}_V B_{ij} = 
B_{kj} \partial_i V^k + B_{ik}\partial_j V^k = 0$.
This is equivalent to the statement that the $1$--form $B_{ij} V^j$ is closed,
or that $V^i = \theta^{ij} \partial_j \rho$. The reparameterization of the world--volume
$\Sigma$ can be equivalently represented by a change in the functions $x^i(\sigma)$ given by
\be\label{eq1000}
x^i \rightarrow x^i - \theta^{jk} \partial_j \rho  \partial_k x^i = x^i + i \{ \rho, x^i
\},
\ee
where we have introduced the Poisson bracket
$$
\{A,B\} = i\ \theta^{ij}\ \partial_i A \partial_j B
$$
on the brane world--volume. Note that
$$
\{ \sigma^i, \sigma^j \} = i\ \theta^{ij},
$$
and more generally that
$$
\{ \sigma^i, A \} = i\ \theta^{ij}\partial_j A.
$$
Written in terms of $\hat A_i$, equation (\ref{eq1000}) reads
\be\label{eq500}
\hat A_i \rightarrow \hat A_i + \partial_i \rho + i\{ \rho, \hat A_i\}.
\ee
Let us mention, for completeness, the following consistency check. Consider, in equation 
(\ref{eq2000}), a variation $\delta a_i (x)$ which is pure gauge $\delta a_i (x) = \partial_i
\lambda(x)$. It is then clear that, if we define $\rho(\sigma) = \lambda(x(\sigma)),$ we have
that $\partial_k \rho(\sigma) = \partial_k x^l \delta a_l(x(\sigma))$. Therefore the variation
of $x^i(\sigma)$ is given by
$$
\delta x^i =\theta^{jk} \partial_j x^i \partial_k \rho = i\{\rho, x^i\}
$$
and is therefore pure gauge.

In the last part of this section we wish to connect the above discussion to that of \cite{SW}.
In \cite{SW} the relation between $a$ and $\hat A$ is derived starting from the knowledge of
the correct non abelian gauge invariance $
\hat A_i \rightarrow \hat A_i + \partial_i \rho + i [ \rho, \hat A_i ]$, where
$[A,B] = A\star B -B\star A = \{A,B\} + o(\theta^2)$. Equivalently, we can start with the gauge
transformation (\ref{eq500}) and follow the argument of \cite{SW} to derive (\ref{eq600}).
In fact, the computation is exactly the same, since all the formulae are identical to 
first non--trivial order in $\theta$. In the treatment in this note the noncommutative
gauge group is the set of diffeomorphisms of the brane world--volume which leave the 
$2$--form $B_{ij}$ invariant. The group is infinitesimally generated by the vector fields
of the form $V^i = \theta^{ij} \partial_j \rho$, which are nothing but the Hamiltonian 
vector fields defined in terms of $B_{ij}$, now considered as a symplectic structure on 
$\Sigma$.

We have worked in the case of $p$ odd. Let me now very briefly discuss 
the case of even $p$, when $B_{ij}$ cannot be invertible. Divide the coordinates 
$X^i$ in $X^0$ and $X^a$, $a=1,\cdots,p$ and assume that $B_{0a} = 0$ and that
$B_{ab}$ is invertible.
Let us consider equation (\ref{eq3000}). For a correct choice of $\lambda$ we can work under
the assumption that $a_0 = 0$. It is then clear that we can solve equation (\ref{eq3000}) by
imposing 
\be\label{eq4000}
d^0 = 0\ \ \ \ \ \ \ \ \ \ x^0(\sigma) = \sigma^0
\ee
and by choosing $d^a = \theta^{ab} a_b$, where $\theta^{ab}$ is the inverse of the invertible
part $B_{ab}$ of the background field strength. It is also clear that we can impose the 
constraint (\ref{eq4000}) not only for small fluctuations of the commutative gauge
potential $a_i$, but, in complete generality, to all orders in 
$a_i$ (this is the choice discussed in \cite{SC}, where
the attention was focused on the case $p=2$ for physical reasons). The constraint (\ref{eq4000})
restricts the noncommutative gauge group to the diffeomorphism which leave the
background field strength and (\ref{eq4000}) invariant. This group is
generated by Hamiltonian vector
fields $V^i$ with $V^0= 0$ and which satisfy $\partial_i \rho = B_{ij} V^j$. This means that
$\rho$ is time independent and that we can essentially reduce the problem by one dimension,
therefore going back to the case previously discussed.

\sect{The $\alpha ' \rightarrow 0$ Limit}

In this last section we analyze the $\alpha ' \rightarrow 0$ limit considered in 
\cite{SW}. Following the reasoning in \cite{SC} and using the results of the last
section we streamline the proof of the equivalence of the 
standard Born--Infeld action with noncommutative Yang--Mills theory
in the large wave--length regime. We will work again for convenience in the 
case $p$ odd. This is done both for notational simplicity and since the case 
$p=2$ was treated in detail in \cite{SC}. In what follows we shall closely
follow the notation of \cite{SW} in order to make quick contact with the results
of that paper.

As in the last section we fix the $U(1)$ field strength to be $B_{ij}$ and consider as
dynamical fields the embedding functions $x^i(\sigma)$. Following \cite{SW}
we assume that the metric in the target spacetime is given by
$$
g_{ij}.
$$
The induced metric on the brane is then given by
$$
h_{ij} = \partial_i X^k \partial_j X^l\,g_{kl}.
$$
Again as in \cite{SW}, we consider the following limit\footnote{The constant $\eta$ in \cite{SC} 
is related to $\epsilon$ in \cite{SW} by $\epsilon = \eta^4$. With this redefinition one can
easily check that the limit considered in \cite{SC} is exactly the same as that analyzed in
\cite{SW}, in the particular case of $p=2$ and ${\rm rk}\ B_{ij} = 2$.} (recall that 
we are considering the case of $B_{ij}$ of maximal rank $p+1$)
$$
g_{ij}\sim \epsilon\ \ \ \ \ \ \ \ 
\alpha ' \sim \epsilon^{1/2} \ \ \ \ \ \ \ \ g_s \sim \epsilon,
$$
where we take $\epsilon\rightarrow 0$.
The tension of the $Dp$--brane is given by $T \propto 1/(g_s l_s^{p+1})$ and therefore scales
as
$$
T \sim \epsilon^{-\frac{p+5}{4}}.
$$
We can then expand the Born--Infeld action as
\beas
S &=& T \int_\Sigma d^{p+1}\sigma \sqrt{\det\left(h_{ij} + 2\pi\alpha ' B_{ij} \right)}\\
&=& T \int_\Sigma d^{p+1}\sigma \sqrt{\det\left(2\pi\alpha ' B_{ij} \right)} \left( 1- 
\frac{1}{4}\left(\frac{1}{2\pi\alpha '}\right)^2 
\theta^{ij} \partial_j X^a \partial_k X^b g_{ab} 
\theta^{kl} \partial_l X^c \partial_i X^d g_{cd}\right)+\cdots\\
&=& {\rm const.}\, -\, \frac{T}{4}  \left(\frac{1}{2\pi\alpha '}\right)^2 
 \int_\Sigma d^{p+1}\sigma \sqrt{\det (2\pi\alpha ' B_{ij})}
\ \{X^a,X^c\} \{X^b,X^d\} g_{ab} g_{cd} + \cdots.
\eeas
The action scales as $T (\alpha ')^{\frac{p-3}{2}} (g_{ab})^2$, and is therefore 
finite in the $\epsilon
\rightarrow 0$ limit, thus showing that the Born--Infeld action does have a smooth polynomial
limit, if it is written in terms of the correct variables. 

To make contact with \cite{SW}, let us define, as in the last section, the noncommutative gauge
potential $\hat A_i$ by the equation
$$
X^i = \sigma^i + \theta^{ij} \hat A_j.
$$
If we define the noncommutative field strength by
$$
\hat F_{ij} = \partial_i \hat A_j - \partial_j \hat A_i - i \{\hat A_i,\hat A_j\}
$$
we can readily check that
$$
-i\,\{ X^i, X^j\} = \theta^{ij} +  \theta^{ik}\theta^{jl} \hat F_{kl}.
$$
Define as in \cite{SW} the open--string metric
\beas
G_{ij} &=& -(2\pi\alpha ')^2 B_{ik} g^{kl} B_{lj} \\
G^{ij} &=& -\frac{1}{(2\pi\alpha ')^2} \theta^{ik} g_{kl} \theta^{lj}
\eeas
and the open--string coupling constant
$$
G_s = g_s {\det}^{1/2}\left( 2\pi\alpha ' B_{ik} g^{kj} \right) \sim \epsilon^\frac{3-p}{4}.
$$
It is then quick to check that 
$$
-\left( \frac{1}{2\pi\alpha '} \right)^2 \{X^a,X^c\}  \{X^b,X^d\} g_{ab} g_{cd} = 
\left({2\pi\alpha '} \right)^2 G^{ab}G^{cd} \hat F_{ac}
 \hat F_{bd} + {\rm const.} + {\rm total\ derivative}.
$$
Therefore the action $S$ becomes
$$
S = {\rm const.}\, + \frac{1}{4 G_{YM}^2} \int_\Sigma d^{p+1}\sigma \sqrt{\det G_{ij}} 
G^{ab}G^{cd} \hat F_{ac} \hat F_{bd},
$$
where
$$
G_{YM}^2 =  \frac{1}{(2\pi)^{p-2} G_s l_s^{p-3}}
$$
is the noncommutative Yang--Mills coupling, which is independent of $\epsilon$.

This then concludes the proof of the equivalence of the commutative and noncommutative
descriptions in the large wave--length approximation. It is uniquely based on the invariance
under diffeomorphisms of the Born--Infeld action. This fact both simplifies the proof 
considerably and also clarifies the geometric nature of the field redefinition to leading
order in the derivative expansion. Let me also note, as a conclusion, that the analysis and
the results in \cite{SW}, restricted to the description of flat $D2$--branes, do give a formal
and complete proof of the conjecture stated in \cite{SC}.

\vs{5}
\noindent
{\bf Acknowledgments}

We would like to thank R. Schiappa for helpful discussion and comments.

\newcommand{\NP}[1]{Nucl.\ Phys.\ {\bf #1}}
\newcommand{\PL}[1]{Phys.\ Lett.\ {\bf #1}}
\newcommand{\CMP}[1]{Comm.\ Math.\ Phys.\ {\bf #1}}
\newcommand{\PR}[1]{Phys.\ Rev.\ {\bf #1}}
\newcommand{\PRL}[1]{Phys.\ Rev.\ Lett.\ {\bf #1}}
\newcommand{\PTP}[1]{Prog.\ Theor.\ Phys.\ {\bf #1}}
\newcommand{\PTPS}[1]{Prog.\ Theor.\ Phys.\ Suppl.\ {\bf #1}}
\newcommand{\MPL}[1]{Mod.\ Phys.\ Lett.\ {\bf #1}}
\newcommand{\IJMP}[1]{Int.\ Jour.\ Mod.\ Phys.\ {\bf #1}}
\newcommand{\JP}[1]{Jour.\ Phys.\ {\bf #1}}
\newcommand{\JMP}[1]{Jour.\ Math.\ Phys.\ {\bf #1}}

\end{document}